# The quantum eraser does not always erase

A. Cardoso[1], J. L. Cordovil[1] and J. R. Croca[1,2]

1 – Centro de Filosofia das Ciências da Universidade de Lisboa, Faculdade de Ciências da Universidade de Lisboa, Campo Grande, Edifício C4, 1749-016 Lisboa, Portugal
2 – Departamento de Física da Faculdade de Ciências da Universidade de Lisboa, Campo Grande, Edifício C8, 1749-016 Lisboa, Portugal

Abstract: In this paper we will first look at a particular quantum eraser setup to show that this type of experiments can be understood in an intuitive manner if we are willing to take a complex nonlinear approach, without the need to invoke Niels Bohr's complementarity or quantum entanglement between two particles. We will then discuss a recent experiment of the same type that does not erase the interference pattern when which-path information is available, and argue that this result is in clear contradiction with the orthodox interpretation of quantum mechanics but perfectly understandable in the framework of nonlinear quantum physics.

## 1. Introduction

The quantum eraser, originally proposed in 1982 by Scully and Druhl[1], has been performed during the last few decades by several authors[2] and using various setups. The main idea in this type of experiments with quantum particles is that the incompatibility between which-way information and interference phenomena can be manipulated using the so-called "entanglement" between two particles. The results have been used as an argument to confirm the orthodox interpretation of quantum mechanics[3] by invoking either Niels Bohr's complementarity or – in some misleading way – Heisenberg's uncertainty principle, depending on the author's preference.

As is well known, if in a simple double-slit experiment a which-path measurement is made, i.e., if we observe which slit the quantum particle goes through, then no interference pattern is observed. That is, it is not possible to have which-way information and an interference pattern at the same time. This can be "explained" by a traditional application of Bohr's complementarity: there are pairs of features (concepts, observables, etc.) which can be observed individually but never at the same time, and both must be used if we want to obtain a full description of a phenomenon. One could, however, argue that such a measurement necessarily disturbs the particle whose state is being observed, and thus it is natural to expect, even without invoking any particular principle, that the interference pattern is simply washed out.

It would therefore be more interesting, as a test to the validity of the orthodox quantum mechanics, if we could find a way to make a which-way measurement without disturbing in any way the object of that measurement, and then check whether we still get an interference pattern. This is precisely what has been achieved three decades ago by Scully and Druhl, who suggested that using an entangled pair of quantum particles, where complementarity has to be applied – as always – to the whole system and not just to its parts or subsystems (the individual particles), it is possible without a local intervention to measure which slit the particle goes through, which as a consequence would eliminate the interference pattern. Moreover, they have argued that by erasing the which-way measurement the interference pattern would be recovered. As mentioned above, these results have been experimentally illustrated using various setups and thus may seem to show, as is argued by many authors, that both complementary and entanglement are real features of quantum systems. The quantum eraser is thus usually presented as a major example of the quantum world's "non-intuitive" character.

In this paper, whose approach is similar to the one in a previous work dedicated to interaction-free measurements[4], we will argue that this is not the case. We will first present and discuss a particular quantum eraser experiment, which employs a double slit to obtain interference, to show that entanglement can be simply understood as a detector-selection effect – without the need to invoke a mysterious interaction between two distant particles – and that this type of experiments can actually be explained in a simple and intuitive manner in the framework of nonlinear quantum physics[5] inspired in the early ideas proposed by de Broglie[6]. We will then discuss a recent double-slit experiment of the same type that does not eliminate the interference pattern even if which-path information is available, and argue that this result is in clear contradiction with orthodox quantum mechanics but is perfectly understandable if we are willing to take a complex nonlinear approach.

## 2. Walborn's quantum eraser experiment

In this section we will focus on a quantum eraser experiment proposed in 1991 by Scully, Englert and Walther[7] and whose scheme was closely exemplified by Walborn, Terra Cunha, Pádua and Monken[8] in 2002. The authors claim their experiment to be the first demonstration of a quantum eraser where interference is obtained using a real double slit. The first part of the experiment can be presented using the setup shown in Fig. 1.

A UV pump beam is injected onto a nonlinear crystal NL, which transforms an incoming photon $\psi$ into a pair of photons – one called idler and represented by the wave-function $\psi_i$ and the other one called signal, $\psi_s$. Only one pair of photons, correlated in space and time and with polarizations perpendicular to each other, is produced at a time.

The idler photon is directly incident onto detector $D_i$ and the signal photon is directed onto a double slit, where each photon's wave-function $\psi_s$ is split into two waves $\psi_{s1}$ and $\psi_{s2}$ with half the amplitude of the original one, such that $\psi_s = \psi_{s1} + \psi_{s2}$. One of them, $\psi_{s1}$, comes out of the upper slit and the other, $\psi_{s2}$, is created at the lower slit. The outcoming waves then diffract and are incident onto detector $D_s$, which is scanned along a direction perpendicular to the signal photon's trajectory and therefore allows for the observation of interference fringes. The signal at $D_s$ is then counted in coincidence with the detections at $D_i$.

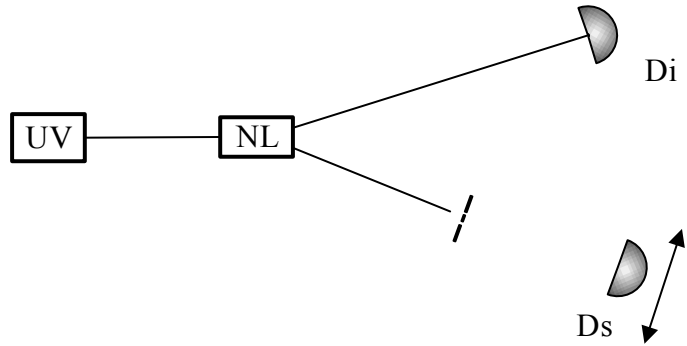

Fig. 1 – Walborn's quantum eraser experiment without polarizers.

In this case it is clear that, as in a simple double-slit experiment, an interference pattern is observed at detector $D_s$.

The authors then place in front of each slit a quarter-wave plate, QW1 at the upper slit and QW2 at the lower one (see Fig. 2), whose fast axes are perpendicular to each other and make an angle of 45° with respect to the signal beam's polarization direction. In this situation the interference pattern previously observed at detector $D_s$ disappears.

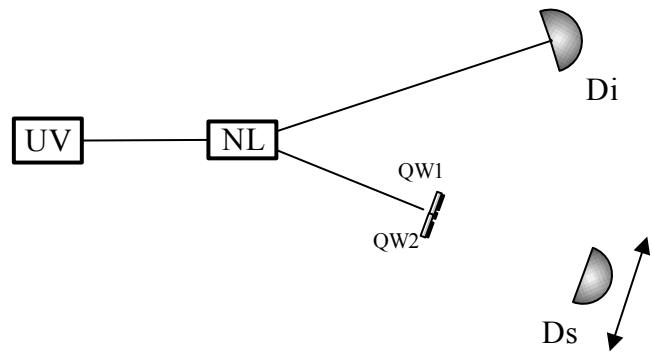

Fig. 2 – Walborn's quantum eraser experiment with quarter-wave plates before the double slit.

Now, as shown in Fig. 3, the authors place after the double slit a linear polarizer cube Pol whose polarization angle is set first as the angle of the fast axis of the quarter-wave plate QW1 and then as the angle of the fast axis of the lower plate QW2. In the first case the interference at $D_s$, whose detections are counted in coincidence with the detections of the idler photons at $D_i$, is recovered in the *fringe* pattern and in the second case it reappears in the *anti-fringe* pattern. The averaged sum of these two interference patterns gives a pattern without interference, as in the previous setup without the polarizer Pol.

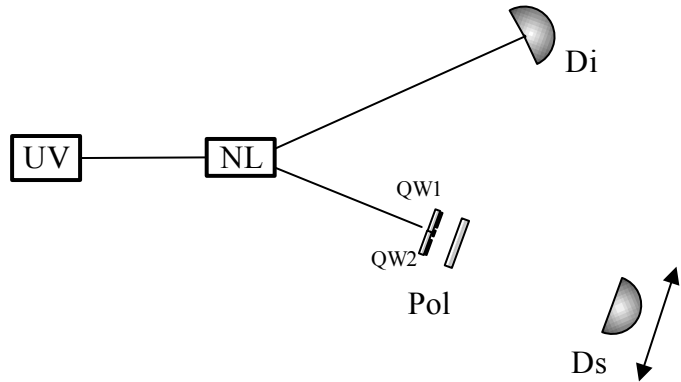

Fig. 3 – Walborn's quantum eraser experiment with quarter-wave plates before the double slit and linear polarizer in the path of the signal photon.

Finally, the polarizer Pol is moved into the path of the idler photon $\psi_i$ (see Fig. 4). This change has no effect on the previous result and the interference pattern at detector $D_s$ in coincidence with the detections at $D_i$ remains the same.

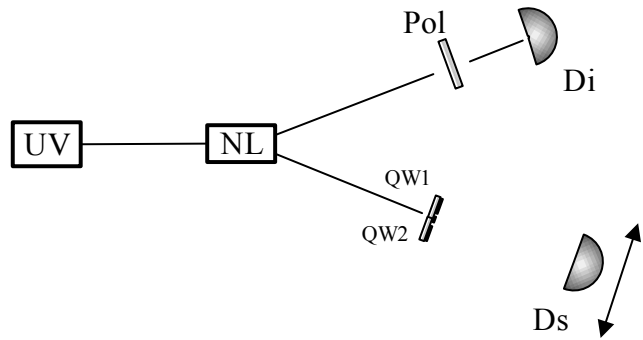

Fig. 4 – Walborn's quantum eraser experiment with quarter-wave plates before the double slit and linear polarizer in the path of the idler photon.

We will now discuss these results in the framework of orthodox quantum mechanics as well as in complex nonlinear terms.

### 2.1. Orthodox interpretation

According to the orthodox interpretation of quantum mechanics, before any measurement is made the signal photon $\psi_s$ that arrives at the double slit is in a linear superposition of two states whose equal probabilities are represented by the waves $\psi_{s1}$ and $\psi_{s2}$.

In the first part of the experiment, before inserting the quarter-wave plates, we have no information about the slit which the photon goes through, and thus the wave-function remains the same, $\psi_s = \psi_{s1} + \psi_{s2}$, after the slits and an interference pattern is observed at

$D_s$.

Now the quarter-wave plates are placed in front of the double slit. As the idler and the signal photons produced at the nonlinear crystal NL have perpendicular polarizations, measuring the polarization of the idler photon reaching detector $D_i$ gives us the polarization of the corresponding signal photon. Therefore, because the latter goes through one of the quarter-wave plates, we are able to know, depending on the direction of the circular polarization with which it arrives at detector $D_s$, which slit the particle went through. In this situation the wave-function $\psi_s$ has to collapse, meaning that $\psi_s \rightarrow \psi_{s1}$ if the signal photon goes through the upper slit and $\psi_s \rightarrow \psi_{s2}$ if it passes through the lower one. Therefore, according to Niels Bohr's complementarity – which states that the wave and particle pictures are complementary and mutually exclusive – the interference pattern disappears.

The polarizer cube Pol is then inserted after the double slit and as a consequence the signal photon's polarization becomes again linear, with an orientation that does not depend on which path the particle went through. This means that the which-way information is lost and thus an interference pattern is recovered at detector $D_s$.

Finally, the linear polarizer is placed in the path of the idler photon $\psi_i$, whose polarization is thus altered to become parallel or perpendicular to the axes of the quarter-wave plates. Consequently, the polarization of the signal photon is altered in the same way due to the entanglement of the photon pair and so its polarization does not change as it goes through either plate QW1 or QW2. This means that, as in the previous situation, we lose the ability to know which slit the particle goes through and thus the interference pattern is recovered at $D_s$.

The discussion above is a clear example of the non-intuitive character of both complementarity and quantum entanglement imposed by this interpretation. In particular, the orthodox approach does not provide us with a physical process through which a change in the polarization of an idler photon results in a change of the polarization of the corresponding signal photon. Also, as mentioned previously for the simple double slit experiment, one could argue that the elimination of the interference pattern in the first place is due not to the which-way information but to the disturbance of the signal photon during its travel to the detector, which in this particular setup is caused by the quarter-wave plates placed before the slits. This makes the use of Bohr's complementarity unnecessary in this case.

## 2.2. Complex nonlinear approach

This non-intuitive character of the quantum world can be easily avoided in the framework of nonlinear quantum physics. In this approach the photon, or any quantum particle, is composed of an extended real wave – the guiding wave, theta wave or subquantum wave – plus a highly energetic and very well localized kernel, corpuscle or acron. This physical wave guides the corpuscle according to the principle of eurhythmy, i.e., preferentially to the regions were the intensity of the wave is greater.

According to this complex nonlinear approach, the real guiding wave $\psi_s$ is split into two waves $\psi_{s1}$ and $\psi_{s2}$ as it arrives at the double slit. The indivisible kernel or acron, however, goes through only one of the slits with equal probability, and thus the wave going through the remaining slit will carry no acron as it moves towards detector $D_s$.

In the first part of the experiment, before inserting the quarter-wave plates, the signal photon's waves $\psi_{s1}$ and $\psi_{s2}$ diffract after the slits and guide their corpuscle preferentially along a path where their resulting intensity is high, thus originating an interference pattern at $D_s$.

Now the quarter-wave plates are placed in front of the double slit. In this situation, as happens with classical electromagnetic waves, the circular polarizations of the two signal beams leaving the slits, obtained after each signal photon passes through both quarter-wave plates QW1 and QW2, will be opposite to each other and consequently there will be no interference at $D_s$.

Then the polarizer cube is inserted after the double slit, selecting only the signal photons whose polarizations are either parallel or perpendicular to the fast axes of the quarter-wave plates they went through before passing the slits, thus eliminating half of the incoming photons for each choice of polarization angle. This means that the polarizations of the signal beams become again linear, as if the quarter-wave plates were not present. In this situation, as only half of the signal reaching detector $D_s$ is counted in coincidence with $D_i$ for each choice of angle at the linear polarizer, the interference pattern is thus recovered.

Finally, the polarizer Pol is moved into the path of the idler photon $\psi_i$. In this case, because the photon pair is correlated in space and time, the results of the coincidence detections will naturally be the same as in the previous setup and an interference pattern will still be obtained at detector $D_s$.

As seen in the previous discussion, our complex nonlinear approach is able to explain the apparent effect of the linear polarizer placed in the path of the idler photon on the behavior of the signal photon detected in coincidence. Moreover, from this point of view it is natural that the averaged sum of the *fringe* and *anti-fringe* interference patterns obtained using the polarizer Pol gives us a total pattern without interference: adding up the detections of the signal photons with polarizations perpendicular to each other allows us to recover the complete signal obtained without the selection made by the polarizer. In sum, all that has been done was to select, by an appropriate choice of the polarization angle, the part of the signal that gives us an interference pattern.

## 3. Menzel's double-slit experiment

We will now present and discuss a similar, more recent two-photon double-slit experiment performed by Menzel, Puhlmann, Heuer and Schleich[9] in 2012 and repeated by the same authors one year later. The setup is shown in Fig. 5.

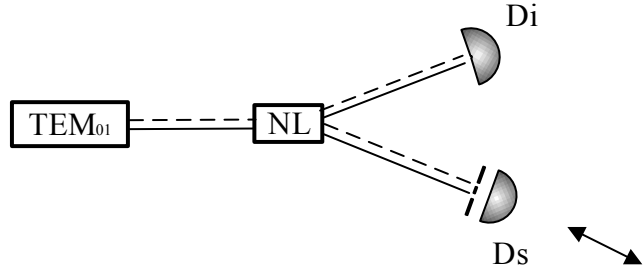

Fig. 5 – Menzel's double-slit experiment.

The UV pump beam employed is now in the $TEM_{01}$ mode, i.e., it is composed of two distinct intensity maxima. Therefore, each photon $\psi$ can be represented by a superposition of two wave functions, one corresponding to the upper maximum, $\psi_1$, and one representing the lower maximum, $\psi_2$. As in the previous experiment, the beam is injected onto a nonlinear crystal NL that can transform an incoming photon into a pair of photons with half the frequency of the first. In this case, however, each produced photon can either be detected at the upper maximum or at the lower maximum, and thus the wave-functions of the idler and signal photons can be written as $\psi_i = \psi_{i1} + \psi_{i2}$ and $\psi_s = \psi_{s1} + \psi_{s2}$, respectively.

The idler beam is again directly incident onto detector $D_i$ and the signal one is directed onto a double slit, in a way that the beam's upper maximum $\psi_{s1}$ is injected onto the upper slit and the lower maximum $\psi_{s2}$ is incident on the lower slit. The outcoming waves then diffract and are incident onto detector $D_s$, which is scanned along a direction perpendicular to the signal photon's trajectory. The signal at $D_s$ can once more be counted in coincidence with the detections at $D_i$.

Now, this experiment is divided in two parts:

1 – Detector $D_s$ is placed just outside the double slit for a near-field detection. In this case, a spatial correlation between the signal and the idler photons is observed: when an idler photon arrives through the upper maximum, which corresponds to the wave $\psi_{i1}$, a signal photon is detected outside the upper slit, $\psi_{s1}$, and if an idler photon arrives through the lower maximum $\psi_{i2}$ then a signal photon is detected outside the lower slit, $\psi_{s2}$.

2 – $D_s$ is now moved away from the double slit for a far-field detection. In this case an interference pattern is always observed when both slits are open, independently on whether the detections at $D_s$ are counted alone or in coincidence with the ones at $D_i$, but disappears when one of the slits is closed.

We will once more discuss the results from the point view of orthodox quantum mechanics and then in terms of a complex nonlinear physics.

## 3.1. Orthodox interpretation

According to the orthodox interpretation, the photon $\psi$ leaving the laser source is in a superposition of two states $\psi_1$ and $\psi_2$ corresponding to the equal probabilities of the particle being detected at the upper and lower maxima, respectively.

When the incoming photon arrives at the nonlinear crystal NL, an outcoming pair of idler $\psi_i$ and signal $\psi_s$ photons is produced in an entangled state. This means that if an idler photon is detected by $D_i$ at the upper maximum then detector $D_s$ will receive a signal photon through the upper slit, which will make its wave-function collapse to the upper state, i.e., $\psi_s \rightarrow \psi_{s1}$, and when an idler photon is detected at the lower maximum a signal photon will be observed through the lower slit, which means that a collapse will occur to the lower state, $\psi_s \rightarrow \psi_{s2}$.

Now, as $D_s$ is moved away from the double slit to allow for the far-field detection, measuring the position of the idler photon gives us the information about which slit the corresponding signal photon goes through. In this situation, as described above, the latter's wave-function will collapse to one of the upper or lower states, and thus $\psi_s \rightarrow \psi_{s1}$ if it comes through the upper slit or $\psi_s \rightarrow \psi_{s2}$ if it arrives through the lower slit. As a consequence, and contrary to the experimental results, no interference pattern should appear at the far screen.

In sum, the which-path information obtained due to the spatial correlation of the photon pair did not collapse the system's wave-function – which had to happen if the wave simply represented the probability of a particle being observed in a certain state – and so did not avoid the observation of an interference pattern. We have therefore shown that, in this particular case, there has been a clear violation of the orthodox interpretation of quantum mechanics, as Bohr's complementarity cannot be invoked in the way that it has been done in previous quantum eraser experiments. Moreover, measuring the state of the idler photon did not have any influence on the signal detections at $D_s$, which means that at least in this case the orthodox notion of entanglement cannot be applied to the photon pair created at the nonlinear crystal.

## 3.2. Complex nonlinear approach

In our complex nonlinear approach, each photon $\psi$ emitted by the UV pump is composed of an acron that is guided by a theta wave with two maxima, an upper one represented by $\psi_1$ and a lower one $\psi_2$. As the acron leaves the laser source along the path of either maxima of its guiding wave $\psi = \psi_1 + \psi_2$, it will tend to remain there due to the principle of eurhythmy, which means that the other maximum will travel alone towards the nonlinear crystal NL.

As the incoming photon enters the crystal, an idler photon $\psi_i$ and a signal photon $\psi_s$ are produced, each one composed of its own acron and guiding wave with the two maxima now represented by $\psi_{i1}$ and $\psi_{i2}$ for the idler photon and by $\psi_{s1}$ and $\psi_{s2}$ for the signal photon.

Moreover, each acron produced will preferentially remain, as it travels along its path, in the same (upper or lower) maximum as the original one.

Now, when they both reach their corresponding detectors in the first part of the experiment, the idler and signal photons will naturally be detected at the same upper or lower maximum, which agrees with the experimental results.

When detector $D_s$ is moved away to the far field, in the second part of the experiment, the signal photon's theta wave $\psi_s$ reaching the double slit will be composed of the same upper $\psi_{s1}$ and lower $\psi_{s2}$ maxima. Therefore, it is obvious that an interference patter will appear independently on whether we detect the signal alone or in coincidence with the idler photons. Naturally, if one of the slits is closed then the wave incident on it will be blocked and thus the interference pattern will disappear.

We thus see that, from our complex nonlinear point of view, it is easy to understand why an interference pattern can be present even if which-path information is available to the observer. Contrary to previous quantum eraser setups, in this particular one the particles going through the double slit have not been disturbed in any way in their path to the detector placed in the far field, and thus there is no reason for the interference pattern to be washed out. Also, as in Walborn's quantum eraser, it is easy with this approach to understand why there is a spatial correlation between the idler and signal photons created at the crystal NL without invoking any mysterious quantum entanglement.

## 4. Conclusions

We have shown in this paper that quantum eraser experiments can be explained in an intuitive manner if we are willing to take a complex nonlinear approach. In particular, from this point of view it is easy to understand why measuring the state of one particle gives us information about the state of another particle sharing a common origin with the first, without the need to invoke a mysterious entanglement between the two particles.

We have also demonstrated that the experimental results obtained by Menzel *et al* are in clear contradiction with the orthodox interpretation of quantum mechanics, as the observation of an interference pattern after a double slit does not depend on whether which-path information is available to the observer. Thus, the incoming wave that splits into two as it reaches the slits cannot simply represent a probability wave.

Finally, we have made it clear why in a quantum eraser experiment like the one performed by Walborn *et al* an interference pattern disappears when which-path information is available but this does not happen in the setup demonstrated by Menzel *et al*. Essentially, a simple and most natural explanation is that there is a real physical wave guiding the particle along its path, and therefore there is no reason for it to collapse even if we know which path the particle goes through.